# On the risk of premature unification

**D Delphenich**
Spring Valley, OH USA 45370

E-mail: feedback@neo-classical-physics.info

**Abstract**: Although the wish to unify theories into something more fundamental is omnipresent and compelling, nonetheless, in a sense, theories must first be "unifiable." The reasons for the success of the unification of electricity and magnetism into a theory of electromagnetism are contrasted with the reasons for the failure of the Einstein-Maxwell unification of gravitation and electromagnetism and the attempts of quantum gravity to unify Einstein's theory of gravity with quantum field theory. The difference between a unification of two theories, a concatenation of them, and the existence of a formal analogy between them is also discussed.

## 1. Introduction

The Holy Grail of all theoretical science is to unify the individual theories for the individual phenomena in terms of some concise set of elementary building blocks that will serve as the basis for all of the specialized theories. That ambition goes back at least as far as the "atomic" hypothesis of Democritus, which amounted to the notion that matter could only be subdivided into smaller pieces a finite number of times before the recursion converged to irreducible units of matter – atoms – that would then form the basic building blocks for all macroscopic matter by essentially reversing the recursion and reassembling the atoms into larger structures.

Interestingly, the point in history at which the atomic hypothesis seemed to have reached its peak of acceptance was perhaps when quantum mechanics succeeded in explaining the periodic table of the elements in terms of the ground state electron shell structure of the atoms. Even at that point in time, the atom was already known to be reducible to a nucleus and electrons. In time, the nucleus was seen to be reducible to protons and neutrons, and eventually the nucleons were thought to be reducible to bound states of quarks. Although that sounds as if a second plateau had been reached in terms of elementarity, in fact, once one takes the view of quantum field theory towards elementary particles as fields that perhaps represent excitations of a quantum vacuum state (which is also a field then, if not a space of fields), one must accept that spaces of fields are typically infinite-dimensional, so one has potentially just opened a new Pandora's box of complexity in the name of elementarity.

Some of the natural tendency towards unification is based in the inductive nature of the scientific method. Typically, one starts with purely empirical observations of new phenomena, which then give way to empirical models (i.e., curve fits to the data), and when one has a large enough number of empirical models that appear to have some common basis, one makes an attempt at a theory that will explain all of them. Eventually, some of the theories seem to exhibit common features, and one attempts to unify the theories into a "meta-theory."

An example of that process is the way that the ancients had noticed that some of the points of light in the sky moved relative to the others and called them "planets." Tycho Brahe amassed a large volume of raw data from his measurements of the line-of-sight angles to the planets as functions of time. Johannes Kepler then organized the data into a set of empirical curve fits that one calls Kepler's laws of planetary motion. Isaac Newton then proposed theories of gravitation and motion that explained not only Kepler's laws of planetary motion, but a large number of other phenomena. Eventually, Einstein expanded the scope of Newtonian gravitation to the strong gravitational fields of dense stellar objects and extended the scope of Newton's laws of motion to relative motions that approached the speed of light.

Other examples of the tendency towards unification will be discussed below.

Before continuing, it is important to clearly distinguish between the unification of theories and their mere concatenation. Basically, a concatenation of theories amounts to a direct product of the spaces involved and the juxtaposition of the known equations for the theories. For instance, one could concatenate Einstein's theory of gravitation with Maxwell's theory of electromagnetism by simply defining the basic field to consist of the Lorentzian metric of the former and the field strength of the latter, while writing down all of the field equations in one place.

However, in order to make that into a true unification of the two field theories, first of all, one would generally expect to have some sort of coupling of the one kind of field to the other, which is referred to as an "induction." Even better, one hopes that there is some general field that would subsume the individual fields in such a way that there would be field equations for that more general field that would reduce to the original field equations in some sort of "classical" limit. Ideally, experimental proof should exist that such inductions actually happen in reality.

Another subtlety that must be addressed is the difference between unifying theories and simply proving that formal analogies exist between them. Typically, the root of all such formal analogies is that one is using the same mathematical techniques to model all of them. For instance, the use of the calculus of exterior differential forms imposes an analogous character on its application to electromagnetism, gravito-electromagnetism, and relativistic hydrodynamics, but it would be premature to suggest that they are unifiable into a common theory of all of them.

That brings us to the topic of the present discussion, namely, the risk of prematurely postulating the unification of theories. In particular, there are some symptoms that perhaps the theories in question are not ready to be unified.

For instance, theories must share a consistent mathematical formalism before they can be easily unified. As we shall see below, part of the reason that the theories of electricity and magnetism could be unified into a theory of electromagnetism was that eventually both of the theories were phrased in the language of vector calculus, in which the formal analogies become more obvious.

Ideally, one should not base a new theory upon other theories, especially when the previous theories are still in a state of being verified experimentally. That defines another form that premature unification can take.

The philosopher of science Karl Popper proposed that given the inductive nature of approaching the truth asymptotically by way of experiments, the only absolute truths of science were essentially the experimental contradictions to its predictions. As Einstein once said of his theory of relativity: "No amount of experimentation can prove my theory correct, but a single experiment can prove it wrong." That would imply that his theory is "falsifiable" in the Popperian sense, and therefore a *scientific* theory, not a pseudoscientific one, by Popper's definition.

Finally, nothing is more convincing of the validity of a new theory than when it implies new phenomena that are ultimately verified experimentally. For instance, a consequence of the Hertz-Maxwell unification of the theories of electricity and magnetism into a theory of electromagnetism was that the inductions between them, one of which (viz., electromagnetic induction) was already known in the laboratory, led to a theory of light and optics in terms of wave-like solutions of the Maxwell equations. Similarly, Einstein's special and general theories of relativity led to numerous predictions of previously-unobserved phenomena that were subsequently observed, such as the contraction of lengths and the bending of light rays when passing gravitating bodies, such as the Sun.

The basic flow of ideas in the rest of this article will be to first examine the theory of electromagnetism as the example *par excellence* of a successful unification of physical theories. We will then examine the unification of space and time by Minkowski as an example of a unification that still has remaining flaws and then discuss the classic problem of early Twentieth-Century theoretical physics that we call the Einstein-Maxwell unification problem. We will then discuss the rise of gauge field theories as the most popular bet on the unification of all field theories, and some of the popular, but experimentally unfounded, theories of how to achieve that unification, such as supersymmetry and quantum gravity.

## 2. The unification of electricity and magnetism

To a primitive culture, the empirical roots of electromagnetism, such as lightning, static electricity in fur, and lodestones, would not seem to suggest any common basis. However, over the centuries, a growing body of experimental results tended to point in that direction. The real quantum leap historically was perhaps the development of electrical circuits, at least in the form of batteries, switches, resistors, and conductors, as well as perhaps Leyden jars to serve as capacitors. That allowed Sir Michael Faraday to perform his experiments on electromagnetic induction, and while employed as a bookbinder, moreover.

The mathematical theory of electrostatics had been developed by Charles Augustin de Coulomb, Pierre-Simon Laplace, and Siméon Denis Poisson into a set of partial differential equations for the electrical field strength vector field **E** or a single partial differential equation for the electrostatic potential $\phi$ from which one can derive it. The mathematical theory of magnetostatics was developed by André Ampère into a set of partial differential equations for either the magnetic field strength vector field **H** or the vector potential **A** from which it could be derived.

Hence, both theories fulfilled the requirement that they could be expressed in a common mathematical formalism, such as vector calculus, which came somewhat later with Josiah Willard Gibbs. In that formalism, the static equations of **E** became:

$$\nabla \times \mathbf{E} = 0, \qquad \nabla \cdot \mathbf{E} = 4\pi\rho, \tag{2.1}$$

if one ignores the formation of electric dipoles in material media; the symbol $\rho$ represents the electric charge density.

The definition of an electrostatic potential function $\phi$ for **E** is:

$$\mathbf{E} = -\nabla\phi. \tag{2.2}$$

The equation for the electrostatic potential $\phi$ then becomes the Poisson equation:

$$\Delta \phi = -4\pi \rho. \tag{2.3}$$

Similarly, the magnetostatic equations for **H** (ignoring the formation of magnetic dipoles in continuous media) take the form:

$$\nabla \times \mathbf{H} = \frac{4\pi}{c} \mathbf{J}, \qquad \nabla \cdot \mathbf{H} = 0, \qquad \nabla \cdot \mathbf{J} = 0, \tag{2.4}$$

in which **J** represents the electric current that is the source of the magnetic field. The last of these equations is an integrability condition for **J** that says not all currents can be the sources of magnetic fields, and is a consequence of the first equation in the set.

If one defines a vector potential **A** for **H** by:

$$\mathbf{H} = \nabla \times \mathbf{A} \tag{2.5}$$

then the equations for **A** then become:

$$\nabla \times (\nabla \times \mathbf{A}) = \frac{4\pi}{c} \mathbf{J}, \qquad \nabla \cdot \mathbf{J} = 0. \tag{2.6}$$

The formal analogy between the equations (2.1) and (2.4) or (2.3) and (2.6) then became a strong inducement to look for a unification of the theories.

As far as inductions between the field theories are concerned, as we pointed out above, Faraday had already shown the existence of electromagnetic induction experimentally; that is, time-varying electric fields could induce time-varying magnetic fields. Heinrich Hertz had also formulated this as a mathematical theory. It can be expressed in the language of vector calculus in the form:

$$\nabla \times \mathbf{E} = -\frac{1}{c} \frac{\partial \mathbf{H}}{\partial t}, \tag{2.7}$$

which is clearly an extended form of the electrostatic equation for $\nabla \times \mathbf{E}$. The derivation of **E** from a potential also includes **A** now:

$$\mathbf{E} = -\nabla \phi - \frac{1}{c} \frac{\partial \mathbf{A}}{\partial t}. \tag{2.8}$$

The capstone of the unification of electricity and magnetism was laid by James Clerk Maxwell [**1**] when he postulated the existence of the opposite process, which one might call "magneto-electric" induction. That will then replace the first of (2.4) with:

$$\nabla \times \mathbf{H} = \frac{1}{c} \frac{\partial \mathbf{E}}{\partial t} + \frac{4\pi}{c} \mathbf{J}, \tag{2.9}$$

while the definition of **A** remains unchanged. The relative sign difference between the time derivatives in (2.7) and (2.9) is an important subtlety of these equations.

As mentioned before, some of the consequences of the unification electricity and magnetism include the fact that it also led to a theory of electromagnetic waves as the basis for optics, in the same way that Newton's theory of motion explained more than just Kepler's laws of planetary motion. That fact is particularly transparent when one expresses the Hertz-Maxwell equations in terms of $\phi$ and **A**, but we shall not stop to discuss that otherwise-known process.

## 3. The unification of space and time

In 1908, Hermann Minkowski delivered an address to the 80$^{th}$ Assembly of German Natural Scientists and Physicians in Cologne that began with the almost-Wagnerian overture ([1]):

> "The views of space and time that I wish to lay before you have sprung from the soil of experimental physics, and therein lies their strength. They are radical. Henceforth, space by itself and time by itself are doomed to fade away as mere shadows, and only a kind of union of the two will preserve an independent reality."

Although it would be nice if more modern scientists were so eloquent in their presentation of their theories, nonetheless, over a century later, it is also clear that Minkowski was being somewhat premature in heralding the end of the era in which one could clearly distinguish between the concepts of time and space.

Let us examine the unification of space and time into space-time. First, one starts with their concatenation into a product manifold $\mathbb{R} \times \Sigma$, where $\mathbb{R}$ is the real number line, which represents time points, and $\Sigma$ is a three-dimensional spatial manifold, such as the vector space $\mathbb{R}^3$ or perhaps a three-dimensional sphere $S^3$. The concatenation $\mathbb{R}^4 = \mathbb{R} \times \mathbb{R}^3$ forms the starting point for Minkowski space.

The reliance of general relativity upon the methods of differential geometry demands that one consider the unification of space and time into a space-time manifold that does not typically have a product structure. That involves introducing non-Euclidian topologies into the modeling problem, as well as non-Euclidian geometries. The business of reversing the unification into "time+space" or "space+time" splittings, which are also called "slicings" and "threadings," is more problematic, and deals with the relativistic extension concept of simultaneity at its most fundamental level [**3**]. It can also raise some issues of whether projective geometry is the proper way to project from a four-dimensional space-time to a three-dimensional space, at least as far as some things, such as four-velocities, are concerned [**4**]. However, although studies have been made of more general "foliations" of the space-time manifold into simultaneity leaves, nonetheless, the existence of solutions to the Cauchy problem of general relativity (i.e., the time evolution of an initial gravitational field) often imposes the "cylindrical" topology of $\mathbb{R} \times \Sigma$, anyway.

The concatenation of space and time becomes a true unification only when one introduces the coupling of the two concepts by way of the algebraic constraint that takes the form of the light cones, in the massless case, and the proper-time hyperboloids, in the massive case. Although the roots of the light cones are clearly in the soil of electromagnetism, nonetheless, they also define a basis for Einstein's theory of gravitation. Hence, although that does not represent a unification in the sense of a common field that includes both fields, nevertheless, it does represent a link between the two theories that unifies them in a different sense of the word.

---

([1]) Reprinted in English translation in the book of collected papers on the principle of relativity [2].

As mentioned above, the consequences of the unification of space and time into space-time include many new things that were verified experimentally, and with increasing precision. Hence, it is tempting to think that the prophecy of Minkowski has come to pass in theoretical physics.

However, to this day, it is still quite misleading to think that time is only a dimension of a four-dimensional manifold that treats time and space equivalently. Even within the context of the theory of relativity, one must still distinguish between time, the dimension, and proper time, the curve parameter for massive matter. Similarly, relativity still thinks of causality as being distinct from time, so that cause-and-effect links between space-time events must be tempered with the limitations imposed by the light-cones.

One might still consider Von Laue's [**5**] objection to Einstein's picture of a space-time manifold as something like a spatially-distributed grid of clocks. Von Laue objected to the idea that time could be defined by anything but real systems in a state of evolution, such as real clocks. Unless there were real clocks at the points of space to define an unambiguous sense of time evolution, it would be physically absurd to associate a time coordinate to a point of space. Of course, one might counter that even a point in the vacuum of deep space is still surrounded by celestial objects in a state of motion or evolution (which might possibly be very incremental processes), which is reminiscent of Mach's principle.

It is in the study of dynamical systems that are more general (or perhaps more specialized) than the concerns of cosmology that one always has to consider other aspects of the eternal enigma of time than its purely geometric ones. For instance, the concept of entropy in thermodynamics is often cited as the basis for the time evolution of thermodynamic systems, such as chemical reactions, in which it defines a preferred direction of evolution for a reaction to take place. When a chemical reaction produces a precipitate that falls out of solution, it is generally "unnatural" to expect it to go back into solution as the initial reactants. Similarly, it is absurd to imagine reversing the processes of oxidation, such as burning and rusting.

## 4. The Einstein-Maxwell unification problem

No sooner had Einstein succeeded in finding a geometric basis for the existence of gravitational fields in space-time than he began to wonder if it would be possible to include the theory of the electromagnetic field into a broader theory that would explain both gravitation and electromagnetism in terms of the same – probably geometric – theory. Since the most definitive theory of gravitation was Einstein's general theory of relativity and the most definitive theory of electromagnetism was due to Maxwell, that is why we shall refer to that unification program as the *Einstein-Maxwell unification problem* ([1]).

One limitation to embarking upon such a quest was that, as of the time in which Einstein began to consider the unification problem, there was no experimental evidence to suggest that inductions between electromagnetism and gravitation actually existed. That situation has prevailed to the present time, moreover. One should note that since the strength of the electromagnetic interaction is many orders of magnitude greater than the strength of the gravitational one, presumably, it would be easier to observe the effects that electromagnetic phenomena induce upon gravitational ones (if they exist) than the effects of the opposite coupling.

---

[1] Some discussions of the various attempts to unify Einstein's theory of gravitation with Maxwell's theory of electromagnetism can be found in the references [**6**].

Perhaps the first attempt at solving the Einstein-Maxwell unification problem was made by Theodore Kaluza in 1921 [7] and refined later by Oskar Klein in 1926 [8] ([1]). It basically amounted to a dimensional extension of a metric field theory that added one more dimension to space-time so that the metric tensor field, which usually contained only gravitational potentials, would also include the electromagnetic potential 1-form in its components. Although the field equation for the five-dimensional metric tensor field that Kaluza and Klein proposed was a natural extension of the Einstein field equations for gravitation, nonetheless, it still did not seem to lead to any inductions between the two fields, and thus amounted to a somewhat-cryptic concatenation of the field theories.

One of the more interesting variations on the Kaluza-Klein theory was based in the fact that since a symmetric 5×5 matrix will have fifteen independent components, and the sum of the components of the four-dimensional metric tensor and the electromagnetic potential 1-form will be fourteen, one of the components of the five-dimensional metric tensor field will still be physically undefined. Oswald Veblen [10] and others (e.g., [11]) suggested that one way to remove that ambiguity was to go to the methods of projective differential geometry, and books on "projective relativity" are being written to this day. (See, especially, [12].)

An attempt by Einstein [13] and Schrödinger [14] led to a similar result (see also Lichnerowicz [9]). In that attempt, the extension of the metric tensor was not dimensional, but an extension from a symmetric, covariant, second-rank tensor field to an asymmetric one. Since such a field will have sixteen independent components, that would include the six that one gets from the electromagnetic field-strength 2-form, which is antisymmetric, and the ten that come from the metric tensor field, which is symmetric. Hence, there is, at least, no stray component to define.

One of the approaches to the unification problem that Einstein considered in the late 1920's and early 1930's was what he was calling "teleparallelism" ([2]). In that theory, the basic field is a global frame field on the space-time manifold. It would also have the right number of independent components to subsume the electromagnetic 2-form and the metric tensor, but his attempts at defining field equations led to unphysical solutions, such as a static distribution of gravitating bodies. Nonetheless, since the question of the parallelizability is unavoidable when one is dealing with differentiable manifolds, the general picture of teleparallelism has endured independently of its use in the Einstein-Maxwell unification problem. Indeed, the original formulation of Einstein's equations in terms of Riemannian geometry can be expressed in an equivalent form using the methods of teleparallelism.

An advance of experimental physics that cast a whole new light on the Einstein-Maxwell unification problem was the observation of "gravito-electromagnetism" in modern satellite experiments [18]. That phenomenon amounts to an extension of the well-known analogy between Coulomb's law of electrostatic interaction and Newton's universal law of gravitation to a theory of weak gravitational fields that includes the possibility that the relative motion of a mass will generate a "gravitomagnetic" field that is analogous to the magnetic field that is generated by the relative motion of a charge. Ultimately, one can use an analogue version of Maxwell's equations to describe weak gravitational fields. Since Einstein's theory of gravitation is essentially a strong-field theory whose effects typically become noticeable only when one is close to a dense stellar object, such as a neutron star or black hole, the analogy between Maxwell's equations and weak-field gravity represents a major paradigm shift in the very definition of the Einstein-Maxwell unification problem. That is, would it not be more natural to unify Einstein's theory of gravitation with whatever the corresponding strong-field theory of electromagnetism would be? Even Hans Thirring [19], who was one of the first to speculate on the possible existence of gravito-electromagnetism, posed essentially that question.

---

([1]) André Lichnerowicz chose to attribute that theory to Pascual Jordan and Yves Thiry in the second part of his book [9] on gravitation and electromagnetism.

([2]) The author has compiled a collection of his own English translations of many of the early papers on teleparallelism in a book [15] that is currently available as a free PDF download from his website: neo-classical-physics.info. He has also written some articles [16, 17] that address the issue from a more modern viewpoint.

Einstein himself had sometimes speculated that one probably could not unify the theories of gravitation and electromagnetism without some sort of contribution from quantum physics. If one thinks of quantum electrodynamics as a first attempt at a strong-field theory of electromagnetism then that would seem reasonable.

It should be pointed out that there have been some promising attempts at formulating a strong-field theory of gravitation that can be expressed in an extended Maxwell form [20]. Furthermore, the basis for those theories is the formulation of Einstein's equations in terms of teleparallelism.

In any event, the Einstein-Maxwell unification problem was probably posed quite prematurely, and one expects that perhaps it is still too soon to pose it properly. Moreover, the question arises of whether it is a valid problem: That is, is it possible that theories are formally analogous, but not unified, except insofar as a path exists between the two?

## 5. Gauge theories as a path to unification

The concept of a gauge field theory goes back to the same period of time in which Einstein and others were trying to develop teleparallelism, and was mostly being developed by Hermann Weyl [21], Vladimir Fock [22], and Dmitri Ivanenko [23]. In its earliest form, it was largely a theory of electromagnetism, in which the Lie group $U(1)$, which describes Euclidian rotations in the plane when that plane is the plane of complex numbers. Since that group is Abelian, one thinks of it as an "Abelian gauge theory."

The aspect of the theory of electromagnetism that gives rise to the concept of a gauge is easiest to explain in the formalism of exterior differential forms. First, one assembled by the components of **E** and **B** into the electromagnetic field strength 2-form *F*:

$$F = c\, dt \wedge E + \#_s \mathbf{B}, \qquad (5.1)$$

in which $E$ is the spatial covector that is dual to **E** by the spatial metric, and $\#_s$ is the spatial duality operator, which makes the components of the spatial 2-form $\#_s\mathbf{B}$ equal to:

$$(\#_s \mathbf{B})_{ij} = \varepsilon_{ijk} B^k. \qquad (5.2)$$

The first set of Maxwell equations (viz., $\nabla \times \mathbf{E} = 0$, $\nabla \cdot \mathbf{B} = 0$) can then be expressed as:

$$d_\wedge F = 0, \qquad (5.3)$$

where $d_\wedge$ is the exterior derivative operator. Since that says that $F$ is, by definition, a *closed* 2-form, by the Poincaré lemma, there should exist an electromagnetic potential 1-form $A$ such that $A = d_\wedge A$, at least locally. However, it will not be unique, since one can add any closed 1-form $\psi$ (so $d_\wedge \psi = 0$) to $A$ and produce another 1-form $A + \psi$ that gives $F$ under exterior differentiation. As long as one is dealing with things locally, one can express any closed 1-form such as $\psi$ in the form $\psi = d\lambda$; i.e., it will be exact. The replacement of $A$ with $A + d\psi$ is referred to as a *gauge transformation of the second kind*. In order to relate that to a "gauge transformation of the first kind," one must go beyond the scope of classical electromagnetism and take a hint from quantum wave mechanics, which would represent a charged particle such as an electron by a complex-valued wave function $\Psi$ (at least, non-relativistically). Since a change of *phase* for the wave function would involve multiplication by an element $e^{i\lambda} \in U(1)$, one assumes, by analogy, that the $\lambda$ in $d\lambda$ is associated with an element of the form $e^{i\lambda}$, as well. Notice that if one does not appeal to wave mechanics then there will be no way of resolving whether $\lambda$ generates an element of $U(1)$ or an element of $\mathbb{R}^+$ (the multiplicative group of positive real numbers; i.e., whether one needs to introduce the $i$ into the exponential.

The research into gauge field theories went into a prolonged hiatus in the early 1930's and was not resurrected until the seminal 1954 paper [**24**] of Chen-Ning Yang and Robert Mills on isotopic spin and isotopic gauge invariance, which extended the scope of gauge field theory to non-Abelian gauge groups, namely, $SU(2)$, which is the gauge group of isotopic spin. Eventually, due to the pioneering work of Ryoyu Utiyama [**25**] and Tai-Tsun Wu, along with Yang [**26**], it emerged that gauge fields had much in common with connection 1-forms that took their values in the Lie algebras of the gauge groups, and that the field strength 2-form that they defined by way of their "exterior covariant derivative" would then be the curvature 2-form of that connection 1-form. The field equations typically take the form of extensions of the Maxwell equations to potential 1-forms and field strength 2-forms that take their values in the Lie algebras of the gauge group.

Other examples of gauge field theories beside electromagnetism include theories of the weak and strong interactions, which relate to the groups $SU(2)$ and $SU(3)$, respectively. However, even though gravitation was the first fundamental interaction to be given a manifestly geometric formulation, strangely enough, a gauge theory of gravitation is still open to much debate with many competing candidates. The basic problem still comes down to finding the most suitable gauge group.

One of the reasons for thinking that gauge field theories are the key to the unification of the theories of fundamental interactions is that the theories of electromagnetism and the weak interaction were unified into the so-called electro-weak theory. That unification was based upon the formulation of both field theories as gauge field theories, so the formalism of the two theories was certainly consistent, and then combining the two gauge groups into the product group $U(1) \times SU(2)$, which is still basically a process of concatenation, and then coupling them by the introduction of a process called "spontaneous symmetry breaking," which had previously been known in the context of statistical physics, and especially the theory of ferromagnetism. One of the characteristic features of spontaneous symmetry breaking is that there will typically be a characteristic energy level at which the phase transition takes place.

Since the Lie group $U(1) \times SU(2)$ is four-dimensional (as a real Lie group), a 1-form that takes its values in the Lie algebra of that Lie group will have four components when one chooses a basis for the Lie algebra. Particle physicists tend to regard individual components of a field as separate particles, and in the electroweak theory the four gauge particles are the photon, which relates to the electromagnetic interaction, as well as the $W^+$, $W^-$, and $Z$ bosons that mediate the weak interactions.

Although the electroweak theory is typically regarded as a unification of the two field theories, it still includes some features that imply that the unification is not entirely complete, such as the fact that one still uses two coupling constants.

Most of the other attempts to unify the fundamental interactions have taken the form of gauge field theories that borrow from the successes of the electroweak theory. Typically, one first looks for a gauge group that will include the gauge groups of the theories to be unified as subgroups. The fields of the larger theory then take their values in vector spaces that carry representations of that larger gauge group; typically, those vector spaces will be tensor products of lower-dimensional vector spaces that are assumed to have various symmetries. The actual unification of the field theories is often assumed to take place at some elevated energy level of interaction at which spontaneous symmetry breaking takes place, which can often raise the question of the falsifiability of the theory when the energy level is far beyond the reach of any realistic experiments, which they often are.

The Grand Unified Theories (GUT's) generally only address the unification of electromagnetism with the weak and strong interactions. They introduce even higher energy scales and often lead to consequences that have yet to be observed experimentally, such as magnetic monopoles and proton decay.

The unification of all four fundamental interactions is referred to as a Theory of Everything (TOE). Most of those theories are based upon the Planck scale as a characteristic scale of units, which we shall discuss in a later section, although we point out that it includes a scale of energy that is so high that it could have possibly existed only in the early Big Bang, if at all.

**6. Supersymmetry**
Something else in fundamental physics that suggests a possible need for unification is the fact that there are still basically two types of elementary particles: fermions and bosons. Elementary fermions are described by wave functions with spin 1/2, such as solutions to the Dirac equation. They include all of the quarks and leptons, which add up to twelve particles. A boson is described by a field with integer spin, and usually takes the form of a gauge field in the standard model, except for the Higgs boson. There are six elementary bosons that are currently known, namely, the aforementioned photon, $W^+$, $W^-$, and $Z$ bosons, one has the gluon that mediates the strong interaction and the Higgs boson. Sometimes, the graviton is also included in this list, which would presumably be the quantum version of a gravitational wave.

Despite this obvious lack of symmetry between fermions and bosons, the desire to represent all elementary particles as gauge particles, and not just the bosons, is so strong that particle theorists were long ago led to speculate upon whether there are a number of missing elementary particles that would make the one-to-one correspondence between elementary fermions and bosons complete, and that speculative symmetry was then referred to as *supersymmetry* (SUSY).

Presumably, the existence of the missing particles is due to the existence of some energy level at which supersymmetry beaks down, although its order of magnitude is not entirely clear. However, the experiments that have been performed at accelerators such as the Large

Hadron Collider have always seemed to suggest that the energy level certainly lies beyond the immediate reach of particle accelerators. Of course, that always gives one an excuse to justify the next generation of accelerators that might reach the proper energy level.

**7. Quantum gravity**

Another path to unification that has attracted considerable attention is what has been called "quantum gravity." Of course, one of the fundamental problems of quantum gravity is defining what the phrase means precisely. That, in turn, reverts to the deeper question of what one means by the adjective "quantum."

The most common interpretation of the phrase "quantum gravity" is the attempt to unify Einstein's theory of gravitation with quantum field theory. The result of such a unification would then be a "Theory of Everything."

One of the first problems that one encounters in one's attempt to unify Einstein's theory of gravitation with the other field theories of quantum physics is that they are described in the languages of two inconsistent formalisms. Basically, Einstein's theory is what is often called a "classical" field theory, although the use of the adjective "classical" is often misleadingly pejorative, as if to suggest that such things are now just quaint anachronisms. What one means by the term "classical field theory" is that one has a field that is defined by a tensor or spinor field, such as the metric tensor field, and a set of partial differential equations that the field must satisfy. One then goes about the business of defining boundary-value problems for the field in the static cases and the Cauchy problem (i.e., initial-value time evolution) in the dynamic ones.

By contrast, due to the highly-speculative nature of the world of quantum phenomena, which generally lie orders of magnitude below the threshold of direct observation, it was decided in the very early years of quantum theory that one would have to suspend one's adherence to the methodology of classical physics and appeal to something more phenomenological. Basically, the early quantum theorists were initially taking a sort of "black box" approach to the modeling of quantum phenomena, in the sense of only looking at the relationship between what went in and what came about, but avoiding the temptation to pose a model for what was inside.

In particular, when Werner Heisenberg and Wolfgang Pauli set about the task of constructing a quantum theory of electromagnetism [**27**], Heisenberg introduced some limiting approximations in order to make things more practicable. For one thing, elementary particles were not presumed to interact by way of forces of interaction, but by the exchange of a particle, such as a photon, in the case of quantum electrodynamics. Hence, although the elementary particles were presumed to be represented by wave functions, there was no attempt made to pose any system of partial differential equations for the fields of the elementary particles that one might call the "field equations of quantum electrodynamics." (The Klein-Gordon and Dirac equations apply to only non-interacting particles.) To be fair, no one had any idea how to go about posing that system of equations, and the popular expectation was that if such equations existed then they would probably be so nonlinear and coupled as to banish all hope of finding useful solutions to the static and dynamic problems.

As a result of that situation, rather than treat the interaction of elementary particles as an example of the Cauchy problem, which would involve propagating the particle wave functions from a finite initial time to a finite final time, the "scattering approximation" was introduced. In that approximation, the initial time is $-\infty$, while the final time is $+\infty$. The time evolution operator then becomes the scattering operator, which takes asymptotic incoming particle states to asymptotic outgoing ones. Because that operator is presumed to be linear, one can then apply the usual methods of quantum theory, such as Fourier transforms, with impunity. The scattering approximation also amounts to assuming that the actual interaction of the particles takes place only "inside the black box," so that the incoming and outgoing particles are assumed to be free (i.e., non-interacting). That also has the effect of defining a characteristic time scale in which the interaction takes place, which sets a lower bound on the time interval that can be resolved in the theory.

Hence, in order to have a better hope of unifying Einstein's theory of gravitation with quantum field theory, one must first make their formalisms consistent. That would involve either finding the elusive "field equations of quantum field theory" (even QED would be a profound advance) or formulating general relativity in the scattering approximation. The latter problem seems unnecessary, since general relativity already devotes considerable attention to the finite-time Cauchy problem, so introducing the approximation would represent a loss of detail in the theory. Similarly, although the experimental observation of gravitational waves would seem to make the concept of a graviton as the quantum gauge particle that mediates the gravitational interaction seem more realistic, nonetheless, one wonders why it would be preferable to say that the reason that the apple fell upon Newton's head is that the apple exchanged a graviton with him.

In the early 1950's, John Archibald Wheeler first proposed the concept of "geons" [**28**] as solutions to his "already-unified" field equations of electromagnetism and gravitation, which amounted to a concatenation of Einstein's equations with Maxwell's equations, when coupled by means of adding the Faraday energy-momentum tensor of the electromagnetic field to the right-hand side of Einstein's equations, which is a concept that goes back to George Yurii Rainich in 1925 [**29**]. In the process, Wheeler had to speculate upon the scale at which one might expect to find geons and introduced a scale of units that actually went back to an Appendix in an 1899 paper [**30**] by Max Planck on the thermodynamics of radiation.

Although Wheeler actually made no mention of that fact is his paper, the scale is currently referred to as the *Planck scale of units*. It is based upon the assumption that one's set of basic physical constants includes only Planck's constant $h$, the speed of light in vacuo $c$, and Newton's gravitational constant $G$. Using only those numbers, one can form a characteristic length, called the *Planck length*, which equals $1.62 \times 10^{-35}$ m, a characteristic time, called the *Planck time*, which equals $5.39 \times 10^{-44}$ s, a characteristic mass, called the *Planck mass*, which equals $2.18 \times 10^{-8}$ kg, and a characteristic energy, called the *Planck energy*, which equals $1.22 \times 10^{19}$ GeV.

Since the highest energy that has been reached by a terrestrial particle accelerator is in the TeV rank ($10^3$ GeV), and the highest energy particles that have been observed by astrophysicists on Earth are ultra-high-energy cosmic rays, which have energies on the order of $10^9$ GeV, one can see that the class of phenomena to which the Planck scale of energy might potentially apply would have to be cosmological in character, such as the very early Big Bang. However, the Big Bang is itself a theory, and it is still in a state of flux regarding the experimental verification of some of its basic assumptions. For instance, the current map of cosmic microwave background radiation, which presumably describes the energy distribution of the universe at an age of about 500,000 years, does not precisely exhibit the spatial homogeneity and isotropy that are basic assumptions of the Standard Model of cosmology, which is based upon the Robertson-Walker solution to the Einstein equations.

Something else that might make the Planck scale a questionable scale of units is that the Planck scale of time is probably orders of magnitude below the scale of time that the scattering approximation imposes upon quantum field theory, which is comparable to the time interval during which a particle interaction takes place. Hence, although some call Planck-scale physics the "physics of extrapolation," nevertheless, it might be better described as the physics of extrapolating beyond the limits of one's approximation. It is analogous to having a power series with a certain radius of convergence and evaluating it for a value of the independent variable lies many light years beyond that radius.

One must admit that the nature of the Planck scale means that any theory about physical phenomena at that scale is not falsifiable. By Popper's definition, that would make it pseudoscientific, not scientific.

Another criticism of Planck scale is that the list of physical constants that were known in 1899 did not include many of the fundamental constants of quantum physics, such as the rest mass of the electron. If one includes that constant, along with the other three, then one can deduce a characteristic scale of length in the form of the Compton wave length of the electron, which is $2.43 \times 10^{-12}$ m and is far better established by experiments, such as electron diffraction. Thus, the argument that the Planck scale is, in any sense, "natural" merely because those are the only characteristic units that can be derived from $h$, $c$, and $G$ is weakened somewhat by the fact that the scale was defined before most of relativity and quantum physics had been developed yet.

Furthermore, many phenomena that quantum gravity regularly deals with, such as wormholes and space-time foam, have not been observed. Considering the popularity of the former concept with the science fiction community, one might suspect that much of modern theoretical physics is drifting into what one might call "science fiction with equations."

## 7. Summary

In summary, we simply say that in order for two theories to be ready for unification, there are certain criteria that should ideally be met:

1. They should be expressed in a common mathematical formalism.

2. They should not be built upon other theories that are still being verified by experiments.

3. There should be some strong indication that couplings (i.e., inductions) might exist between them.

4. They should be falsifiable; i.e., it should be realistically possible to configure the experiments that might confirm or deny their predictions.

__________